\documentclass{article}
\usepackage[margin=1in]{geometry}
\usepackage{authblk}      
\usepackage{graphicx}     
\usepackage{amsmath, amssymb}  
\usepackage{hyperref}     
\usepackage{cite}       
\usepackage{makecell}
\usepackage{natbib}
\defcitealias{us2023using}{FDA, 2023}

\title{Revolutionizing Clinical Trials: A Manifesto for AI-Driven Transformation}
\author[]{}
\date{}  

\begin{document}

\maketitle


\begin{center}
\renewcommand{\arraystretch}{1.5}
\begin{tabular}{ccc}

\makecell[c]{Mihaela van der Schaar\textsuperscript{1} \\ \footnotesize\texttt{mv472@cam.ac.uk}} &
\makecell[c]{Richard Peck\textsuperscript{1} \\ \footnotesize\texttt{richard.peck@liverpool.ac.uk}} &
\makecell[c]{Eoin McKinney\textsuperscript{1} \\ \footnotesize\texttt{efm30@medschl.cam.ac.uk}} \\

\\

\makecell[c]{Jim Weatherall\textsuperscript{2} \\ \footnotesize\texttt{James.Weatherall@astrazeneca.com}} &
\makecell[c]{Stuart Bailey\textsuperscript{3} \\ \footnotesize\texttt{stu.bailey@emdserono.com}} &
\makecell[c]{Justine Rochon\textsuperscript{4} \\ \footnotesize\texttt{justine.rochon@boehringer-ingelheim.com}} \\

\\

\makecell[c]{Chris Anagnostopoulos\textsuperscript{5} \\ \footnotesize\texttt{chris.anagnostopoulos@quantumblack.com}} &
\makecell[c]{Pierre Marquet\textsuperscript{6} \\ \footnotesize\texttt{pierre.marquet@unilim.fr}} &
\makecell[c]{Anthony Wood\textsuperscript{7} \\ \footnotesize\texttt{tony.x.wood@gsk.com}} \\

\\

\makecell[c]{Nicky Best\textsuperscript{7} \\ \footnotesize\texttt{nicky.x.best@gsk.com}} &
\makecell[c]{Harry Amad\textsuperscript{1} \\ \footnotesize\texttt{hmka3@cam.ac.uk}} &
\makecell[c]{Julianna Piskorz\textsuperscript{1} \\ \footnotesize\texttt{jp2048@cam.ac.uk}} \\

\\

\makecell[c]{Krzysztof Kacprzyk\textsuperscript{1} \\ \footnotesize\texttt{kk751@cam.ac.uk}} &
\makecell[c]{Rafik Salama\textsuperscript{8} \\ \footnotesize\texttt{rafik.salama@accenture.com}} &
\makecell[c]{Christina Gunther\textsuperscript{8} \\ \footnotesize\texttt{christina.gunther@accenture.com}} \\

\\

\makecell[c]{Francesca Frau\textsuperscript{9} \\ \footnotesize\texttt{Francesca.Frau@sanofi.com}} &
\makecell[c]{Antoine Pugeat\textsuperscript{9} \\ \footnotesize\texttt{Antoine.Pugeat@sanofi.com}} &
\makecell[c]{Ramon Hernandez\textsuperscript{9} \\ \footnotesize\texttt{Ramon.Hernandez@sanofi.com}} \\

\end{tabular}
\end{center}

\vspace{1em}

\begin{center}
\textsuperscript{1}University of Cambridge \quad
\textsuperscript{2}AstraZeneca \quad
\textsuperscript{3}EMD Serono \quad
\textsuperscript{4}Boehringer Ingelheim \\
\textsuperscript{5}QuantumBlack \quad
\textsuperscript{6}University of Limoges \quad
\textsuperscript{7}GSK \quad
\textsuperscript{8}Accenture \quad
\textsuperscript{9}Sanofi
\end{center}

\vspace{20pt}

\section{Introduction: Transforming Clinical Trials to Improve Patient Outcomes}

Clinical trials are the bedrock of medical practice. They provide a scientifically rigorous way to test the safety and efficacy of new treatments, drugs, and medical devices. Considering the substantial investment required for clinical trials, with Phase III trials often exceeding \$500 million and lasting several years \citep{sertkaya2024costs}, it is crucial to conduct them with utmost efficiency. Furthermore, they represent a cornerstone in the pharmaceutical industry’s ongoing commitment to enhancing patient care strategies. To inform evidence-based practice, trials should accurately represent the diversity and complexity of real-world patient populations, thereby strengthening the evidence for new treatments. However, trials typically exclude $> 75\%$ of patients from testing \citep{he2020exclusion}. Even though treatments are often much more widely used \citep{martin2004differences}. With recent advances in Artificial Intelligence (AI) transforming fields ranging from finance to education, it is essential to explore how these technologies can be harnessed to maximize the potential of clinical trials by improving decision making (earlier termination of unsafe or ineffective treatments, acceleration of safe and effective medicines to market) while driving down costs.

This manifesto represents a collaborative vision forged by leaders in pharmaceuticals, consulting firms, clinical research, and AI. It outlines a roadmap for two AI technologies—causal inference and digital twins—to transform clinical trials, delivering faster, safer, and more personalized outcomes for patients. By focusing on actionable integration within existing regulatory frameworks, we propose a way forward to revolutionize clinical research and redefine the gold standard for clinical trials using AI.

\section{The Goals of AI-Enabled Clinical Trials}
In this section, we outline the key goals of AI-enabled Clinical Trials as identified in the summit. The goals outlined below are by no means exhaustive or complete, but rather aim at highlighting the potential for causal inference and digital twins, specifically, to positively impact trials.

\paragraph{Accelerating Answers.} Traditional trials and drug development programs require a lot of time to demonstrate the benefit-risk profile of new treatments, identify target populations in whom this profile is sufficiently positive, and optimize the dose. They also face challenges with recruitment, adherence, and participant retention. AI can dramatically accelerate the generation of clinical insights and improve their precision by impacting each of these areas \citepalias{us2023using}. Causal inference identifies treatment responders with high specificity, enabling trials to target subpopulations most likely to benefit. Digital twins model individual patient trajectories, predicting safety and efficacy in real time and facilitating within or between trial decision making and supporting effective and efficient adaptive trial designs. For example, in oncology trials, causal models could identify biomarkers for early responders to novel immunotherapies, while digital twins can simulate personalized treatment paths to optimize dose and reduce toxicity. These technologies can ensure that trials answer critical questions faster and/or more effectively, guiding life-saving therapies to the right patients and with greater urgency.

\paragraph{Improving vs. Increasing Probability of Success (POS).} Improving the probability of success is the single most effective way to improve pharmaceutical R\&D efficiency (Paul et al 2010). Not all efforts will increase POS, but overall improvement of POS will lead to better earlier decision making. Causal inference can play a key role in evaluating the likelihood of true benefit rather than random-high bias (add REF) and better contextualize early study results. Increasing the early termination rate of ineffective or unsafe treatments is one key to resource efficiency across drug development. 
For those treatments with evidence of promise, expanded use of causal inference and digital twins to identify the patients or subpopulations most likely to respond, along with the optimal treatment regimens that will improve the benefit-risk profile vs. standard of care medicines, will increase the probability of success in late development. Improving early decision making and increasing late-phase POS will significantly decrease costs, both by reducing the number of negative Phase II or III trials while improving the design of Phase III trials focused on the right populations and/or treatment regimens.

\paragraph{Expanding the Questions Trials Can Answer.} AI unlocks the potential to address previously intractable questions. Current trials often focus narrowly on predefined endpoints, missing opportunities to explore secondary effects, combination therapies, or diverse population responses. By integrating real-world data and advanced AI methodologies, in silico trials can uncover novel insights about disease mechanisms, treatment interactions, and long-term outcomes \citep{pappalardo2019silico}. Digital twins, for instance, can simulate the impact of an anti-diabetic drug on patients with concurrent cardiovascular disease, identifying risks or synergies that conventional methods would overlook. Similarly, causal inference can explore how treatments interact with lifestyle factors like diet or exercise, providing a more holistic understanding of patient care. A better understanding of secondary effects or temporal influences on benefit-risk offers the potential to improve drug labeling and better inform patient care decisions when the drug is on the market, outside of the strictly controlled clinical trial paradigm. 

\paragraph{Balancing Confirmation and Discovery.} AI-driven methodologies can make trials more informative by incorporating diverse data streams. For example, causal inference allows trials to estimate drug efficacy across different demographic groups, ensuring treatments work equitably. Digital twins create synthetic control arms, reducing patient burden while enabling trials to answer more questions simultaneously. The dual purpose of clinical trials—to confirm efficacy and enable discovery—requires a careful balance. While confirmatory trials focus on regulatory endpoints, exploratory trials can leverage AI to uncover new hypotheses, design better protocols, and extract maximum value from existing data. The concept of “Integrated Evidence” is designed to support not just approval by regulatory agencies, but also additional evidence needs for stakeholders such as payers, physicians, patients, and caregivers, aiming to support a wider generation and use of data in both the clinical trials and real-world settings. Currently, a new drug approval by a regulatory agency usually requires two significant pivotal trials: efficacy and safety confirmation trials. However, the introduction of digital twins that can simulate entire clinical trials and predict whole trial outcomes could potentially make replications of pivotal trials obsolete (see Appendix 1). Additionally, AI-driven methodology can be used to design late-phase trials with both confirmatory and subsequent exploratory (optimization) stages <ref>. Ultimately, these advancements could accelerate the delivery of effective treatments to patients.

\section{Digital Twins: Transforming Clinical Trials for Patient-Centered Outcomes}
Digital twins can enable transformative change in clinical trial design and execution, allowing a dynamic, patient-first approach that emphasizes personalization, safety, and diversity. These computational models simulate the biological and therapeutic responses of individual patients or populations, enabling real-time predictions and personalized treatment decision-making. In doing so, digital twins bring clinical trials closer to achieving the promise of precision medicine.

\paragraph{AI-Enabled Digital Twins.} Digital twins have traditionally been constructed using mechanistic models derived from established physiological and clinical systems knowledge \citep{laubenbacher2021using}. These models, while valuable, rely on predefined equations and assumptions that limit their adaptability and precision when dealing with complex, individualized systems. Such approaches struggle to account for the variability and complexity of real-world patient populations, particularly when incorporating dynamic interactions such as comorbidities, lifestyle factors, or evolving health conditions.

AI-driven methodologies, by contrast, are revolutionizing the way digital twins are built and deployed \citep{holt2024automatically}. Using advanced machine learning techniques, such as neural networks, symbolic regression, and differential equation discovery, AI enables the creation of data-driven, adaptive digital twins that evolve in real time. These models can integrate domain knowledge with insights derived from diverse datasets, including electronic health records (EHRs), genomics, longitudinal health data, and wearable device outputs, to construct highly personalized representations of individual patients. AI-based digital twins excel in capturing complex relationships and interdependencies between variables, and even rare or emergent patterns that traditional models might miss. This capability enhances simulation accuracy and enables the simulation of a diverse array of `what-if' counterfactual scenarios, such as testing the efficacy of a range of dosages or simulating responses to different treatment combinations. This also allows for the exploration of different outcomes under varying conditions, which can be crucial in understanding and predicting patient responses.

By moving beyond mechanistic frameworks, AI empowers digital twins to become dynamic tools for precision medicine, capable of adapting to individual needs and clinical contexts. While the potential applications of digital twins throughout the clinical trial process are vast, during the summit, three use cases were given particular focus, which we highlight now.

\paragraph{Enhancing Trial Diversity and Safety using Digital Twins.} In Phase I trials, it is crucial to maintain precision in pharmacovigilance. Stratification of patients into different risk groups via AI/ML can revolutionize the individual safety experience by simulating how patients might react to various dosages. This predictive capability allows researchers to identify potential risk factors for serious adverse effects before even administering a drug, sparing participants unnecessary harm. For example, deep learning methods (neural ODEs) can learn from existing dose regimens to predict PK responses from untested regimens \citep{lu2021neural}. When digital twins are used in the stratification process, they simulate detailed patient reactions to treatments, which is particularly useful for rare conditions or complex cases. For example, in cardiovascular medicine, integration of electrophysiological and anatomical (MRI) data into a twin has been used to predict the occurrence of potentially fatal changes in heart rhythm after heart attacks \citep{arevalo2016arrhythmia}. Twins have also been used to estimate response to pacemaker implantation, facilitating targeted recruitment into trials \citep{viola2023gpu}.

In early clinical trial phases, safety is paramount, as participants may experience unexpected adverse reactions to the novel intervention, since treatment protocols, such as dosing strategies, are not yet fully established. Digital twins can revolutionize safety monitoring in this context by simulating likely participant-drug interactions either before or during the trial, enabling early identification of potentially harmful patient trajectories. If the digital twin of a participant predicts a trend towards an undesirable state, pre-emptive actions can be taken, such as increasing monitoring, adjusting the dosage, or switching treatments, to mitigate harm. For instance, a digital twin of a patient with a metabolic disorder may predict that a standard dose of an experimental drug might lead to toxic accumulation, prompting dose adjustments or additional safeguards before administration. Such intelligent treatment adjustments guided by digital twins can reduce drop-outs due to adverse reactions, which are a significant burden on clinical trials currently \citep{bower2014interventions}. Furthermore, such adjustments enable a greater diversity of patients to be included in the clinical trial, as appropriate treatment adjustments based on the digital twin predictions can ensure the safety of participants who may currently be unnecessarily excluded based on rigid inclusion criteria.
Additionally, even if some trial drop-outs cannot be prevented, digital twins can play a pivotal role in making full use of the data already collected from such participants. If a participant withdraws from a trial, their digital twin could serve as their proxy, continuing to simulate their state based on their historical data and the data of related participants. This allows researchers to make the most of existing data, mitigating losses to statistical power and accumulation of attrition bias from drop-outs. By leveraging digital twins in this triple role—safety enhancement, increasing diversity, and data preservation—clinical trials can become both safer and more resilient to the challenges of real-world drug application.
Finally, we note that while digital twins offer a high level of precision and personalization, even without them, stratification algorithms leveraging AI/ML techniques can still provide valuable insights and improve patient safety in Phase I trials.

\paragraph{Increasing the Efficiency and Diversity of Clinical Trials with Digital-Twin-Based Control Arms across Various Comparators.} Control groups can involve a large proportion of the participants in a traditional trial, which can result in an excessive financial burden in running trials \citep{moore2018estimated}. Furthermore, patients may be unwilling to be randomized into control groups that offer little chance of providing clinical benefit \citep{kadam2016challenges}. They may instead look for other, open-label trials or try treatments outside a trial framework. Therefore, maximizing the proportion of trial participants that can be assigned to novel treatment arms can improve recruitment and increase the potential benefits for participants in such scenarios. The integration of digital-twin-based virtual control arms into clinical trials has the potential to revolutionize trial efficiency and participant experience in this respect, by maximizing the proportion of real patients assigned to novel treatment.
Digital twins enable partial virtual control groups, whereby a minimum number of real patients are assigned to the control group and are supplemented by virtual controls derived from digital twins. Indeed, digital twins can successfully model counterfactual treatment outcomes to such an extent that they can recreate clinical trial results from observational data alone \citep{qian2021synctwin}, highlighting that control groups can be plausibly augmented with accurate digital twin-based simulations of counterfactual outcomes from treated patients. In addition, given this increased efficiency using partly virtual control arms, digital twins can enable trials to incorporate multiple comparator arms, enhancing the relevance and generalizability of results and minimizing the cost of clinical evidence \citep{park2022economic}. Traditional trials typically choose comparators based on regulatory requirements, which may not reflect the most clinically or regionally relevant treatments. For instance, treatment protocols can differ significantly across regions or countries or the clinical standard of care may have changed between completion of the trial and approval of the treatment and/or its adoption into treatment guidelines, and using digital twins to simulate a diverse set of comparators ensures that trial outcomes remain applicable across varied healthcare settings. This approach enhances the global applicability of trial findings, addressing disparities in treatment standards and regulatory expectations over time.

\paragraph{Personalising Treatment Plans to Improve Results in Post-Trial Clinical Deployment.} In the post-marketing surveillance phase, incorporating novel treatments into real-world settings and updating existing treatment protocols can be a challenging process. Digital twins can allow clinicians to simulate patient responses to an array of potential treatments, including newly approved treatments. In doing so, clinicians can increase their understanding of the effects of novel treatments, allowing them to correctly place them among existing interventions and enabling precise personalization of treatment plans. In this way, digital twins can complement any guidelines that stem from the results of a clinical trial to establish personalised treatment protocols, such as establishing precise dosages and treatment timings for individual patients, even without invasive data collection \citep{kuang2024med}. For example, in settings where multiple treatments can be beneficial, deciding the optimal treatment combination, and whether to administer concurrent or sequential treatment, can be non-trivial, e.g., determining the timing of chemo and radiotherapy when treating oropharyngeal carcinoma \citep{cooper2004postoperative, pignon2009metaanalysis}. Simulations with digital twins can provide principled platforms for decision-support in such cases, leading to highly optimal treatment decisions \citep{tardini2022optimal}, which can be especially beneficial when dealing with novel treatments with yet unspecified optimal policies.
Furthermore, given appropriate continuous updating of digital twin models, as novel treatments are exposed to increasingly diverse populations and over longer-term periods during post-marketing surveillance, digital twin-based simulations will become increasingly accurate for novel interventions. This can help ensure that treatment protocols continuously evolve alongside the most recent evidence of treatment efficacy, to assist the decision-making of clinicians in today’s increasingly complex medical environment. 

\paragraph{Summary: Digital Twins as Catalysts for Change.} By simulating and monitoring patient responses, digital twins have the potential to fundamentally reshape the clinical trial landscape. They enable trials that are faster, safer, and more patient-centric, addressing long-standing challenges in efficiency, diversity, and real-world applicability. From predicting individual safety profiles to optimizing treatment regimens and expanding the scope of trial insights, digital twins bridge the gap between clinical research and real-world care.
When combined with other AI-driven methodologies, such as causal inference, which is described next, digital twins offer a powerful, complementary approach to modernizing clinical trials. Together, they represent a new frontier in patient-centered innovation, ensuring that the promise of precision medicine becomes a reality for diverse populations worldwide.

\section{Causal Inference and Discovery: Transforming Clinical Trials into Engines of Insight}

Machine learning (ML)-based causal inference methods are poised to revolutionize clinical trials by ensuring the generation of actionable and reliable insights. These methods can accurately identify true treatment effects while accounting for confounding variables and selection biases, thereby enhancing the validity and robustness of trial outcomes. When integrated with AI-driven discovery tools, such as symbolic regression and differential equation discovery, causal inference not only optimizes trial designs but also uncovers fundamental scientific knowledge, extending the impact of clinical research beyond traditional boundaries. Moreover, ML-based causal inference can leverage large-scale and rich datasets – including electronic health records (EHRs), biobank repositories, and data from failed or negative trials – to systematically extract valuable information that would otherwise remain untapped. This integrative capacity broadens the evidence base beyond traditional randomized controlled trials (RCTs), improving external validity and ensuring that findings are generalizable to more diverse real-world populations. Ultimately, this data-driven paradigm allows for more efficient trial design, better targeting of scarce resources, and an accelerated path from discovery to clinical translation.

\paragraph{Identifying Predictive Biomarkers Through ML-Empowered Causal Inference.} Traditionally, decisions about which biomarkers to measure in a clinical trial follow expert-led hypotheses drawn from domain knowledge, mechanistic studies or limited early-phase results. This approach can overlook subtle but clinically important variables and often fails to account for confounding or complex covariate interactions. Modern ML-empowered causal inference methods, however, focus on heterogeneous treatment effect (HTE) estimation to directly quantify how specific biomarkers influence treatment outcomes, moving beyond simplistic correlations and ensuring that any identified biomarker is genuinely predictive of heterogeneity in treatment effect.
 
By leveraging large-scale and rich datasets, such as electronic health records (EHRs), biobank repositories, or data from failed and negative trials, ML-empowered causal inference methods can systematically screen thousands of candidate variables, accounting for complex patient histories and a broad range of covariates. Advanced techniques like causal forests, doubly robust estimators, or neural network-based representation learning incorporate sophisticated regularization and variable selection, uncovering nonlinearities and gene-environment or biomarker-drug interactions that simpler methods routinely miss \citep{chernozhukov2016double, van2011targeted, shi2019adapting}. As a result, teams can isolate biomarkers that meaningfully shift the HTE, providing a principled rationale for their inclusion in prospective studies. For example, consider planning an oncology trial targeting metastatic melanoma. Applying ML-based causal inference to pre-existing EHR data might reveal that a previously underappreciated inflammatory biomarker is not only correlated with improved response but actually modulates the effect of immunotherapy. Armed with this HTE-driven insight, sponsors can prioritize measuring that biomarker in the next trial, refine their eligibility criteria, and implement adaptive dosing strategies. The outcome: more efficient enrollment, more targeted interventions, and ultimately, accelerated drug development that resonates with both ML methodologists and pharmaceutical stakeholders.

\paragraph{Subgroup Analysis for Tailored Treatments Through CATE Estimation.} While the identification of relevant biomarkers ensures that clinical trials measure the right variables to capture treatment-effect heterogeneity, it is equally important to determine which patient subgroups stand to benefit most. Traditional analyses of RCT data typically produce a single average treatment effect (ATE), collapsing heterogeneous responses into a single, often overly generic measure. In contrast, conditional average treatment effect (CATE) estimation, facilitated by ML methods, characterizes a continuous spectrum of treatment responses across varying patient characteristics. By shifting from a single-point summary to a rich treatment-response surface, CATE estimation supports a data-driven approach to trial design – one that pinpoints groups with particularly favorable or unfavorable responses and informs more individualized treatment or dosing recommendations \cite{bica2021real}, thus paving the way towards precision medicine.
Crucially, CATE estimation does not need to rely solely on data from a single RCT. By incorporating well-curated observational data—such as EHRs, post-marketing surveillance studies, or disease registries—and applying rigorous causal inference methods that adjust for confounding (e.g., inverse probability weighting, targeted maximum likelihood estimation), researchers can leverage large and diverse patient populations to refine their subgroup definitions. For example, consider developing a new antihypertensive therapy. Beyond simply knowing which biomarkers to measure, sponsors can combine their RCT dataset with real-world EHR data to identify not just a single ``optimal'' biomarker-defined stratum, but a more nuanced set of patient profiles – such as those with both hypertension and mild chronic kidney disease – who display especially pronounced improvements.  This integrated, data-rich approach ensures that the subsequent trials can be more sample-efficient, targeting recruitment towards these promising subgroups from the start, and ultimately leading to more personalized, effective interventions that resonate with both methodologists and pharmaceutical stakeholders \cite{athey2019estimating}.

\paragraph{Generalizing Trial Findings to Real-World Populations.} While identifying relevant biomarkers and tailored subgroups helps clinical trials produce more targeted and internally valid results, it does not guarantee that these findings will hold beyond the controlled trial environment. Real-world populations often differ from trial cohorts in demographic makeup, clinical practice patterns, and comorbidity profiles. ML-based causal inference methods enable the transport of trial-based estimates to these new settings by leveraging large observational datasets—such as EHRs, registries, and post-marketing surveillance studies—and applying transportability analyses and robust adjustments for confounding to align trial populations with broader patient groups \citep{bareinboim2016causal, pmlr-v235-schweisthal24a}. By properly aligning the trial population with the target setting, these methods yield effect estimates that better reflect routine clinical environments. For instance, consider a novel heart failure therapy tested primarily in relatively young, low-risk patients. ML-driven causal transport methods can integrate EHR data from older, multimorbid patient populations -- where real-world medication adherence, polypharmacy, and healthcare system variability come into play—to refine the expected benefit-risk ratio of the therapy in these more complex scenarios. By revealing how both effectiveness and safety metrics shift under different care settings, these ML-based approaches guide more informed decision-making for clinicians, as well as regulators. Ultimately, this data-driven strategy ensures that treatments proven effective and safe in tightly controlled trials can be responsibly extended to the full spectrum of patients who could benefit, bolstering confidence in new therapies and fostering safer, more equitable healthcare delivery at scale.

\paragraph{Discovery-Driven Insights from Clinical Trial Data.} AI-driven discovery methods, integrated with causal inference, unlock the latent potential within clinical trial data to uncover new mechanisms and improve understanding of drug behavior. Traditional physiological and PKPD models rely on predefined structures that often fail to capture patient-level variability or novel interactions. Machine learning techniques like symbolic regression \citep{bongard2007automated} and ordinary/partial differential equation (ODE/PDE) discovery \citep{brunton2016discovering} overcome these limitations by creating data-driven models that reveal unexpected relationships. For instance, in a trial for a novel anti-inflammatory drug, symbolic regression might uncover an interaction between body mass index, genetic markers, and drug metabolism, identifying new biomarkers for efficacy. Differential equation discovery could then simulate how these biomarkers evolve over time, providing insights into the drug’s long-term effects and informing subsequent trial designs.
Personalized ODEs \citep{kacprzyk2024ode}, which bridge traditional ODE modeling with treatment effect estimation, represent a significant breakthrough in individualized modeling. These models accommodate variability across patient populations, improving accuracy and enabling tailored treatments. For example, in a trial for a new antihypertensive drug, personalized ODEs could model how an individual’s blood pressure responds to different dosing regimens over time, optimizing both efficacy and safety.

\paragraph{Summary: Causal Inference and Discovery as Pillars of Modern Clinical Trials.} Causal inference and AI-driven discovery represent a paradigm shift in clinical trial methodology, enabling precision, scalability, and deep scientific insight. By enabling data-driven biomarker discovery, heterogeneous treatment effect estimation, and rigorous generalization to real-world populations, causal models ensure trials generate actionable insights with tangible benefits for patients. Meanwhile, discovery tools like symbolic regression and differential equation modeling uncover novel mechanisms and relationships, advancing our understanding of drug behavior and unlocking the full potential of trial data.
Together, these approaches transform clinical trials into platforms for both validation and discovery, bridging rigorous scientific inquiry with real-world patient care. Their integration into clinical research promises not only safer and more effective therapies but also groundbreaking insights that shape the future of medicine. As we move forward, these technologies will be indispensable in driving innovation and delivering on the promise of precision medicine for diverse patient populations worldwide.

\section{A Framework for Robust Validation of AI Methods in Clinical Trials}

Rigorous validation of AI-driven methodologies is necessary to ensure reliability, generalizability, and alignment with regulatory standards. Given the significance of this, validation practices were a large focus of the summit. Currently, there is a lack of standardized approaches to validation, leading to inevitable confusion about whether AI methods are suitably mature and effective to adopt in clinical settings. We distil the summit’s discussions into a proposed validation framework below (summary in Table \ref{tab:validation-framework}), outlining critical components to ensure robust validation of AI methods for clinical trials, and their relevance to digital twins and causal inference.

\paragraph{Define the Context of Use.} The first step in any validation process must be to define the precise context in which the AI model will operate within the clinical trial. For digital twins, this could involve their use in simulating patient-specific outcomes for dose optimization, while for causal inference models, the use might be identifying treatment effects across specific subpopulations. Clear articulation of objectives ensures that the resulting validation strategy is tailored to the task at hand. For example, the above use cases should guide how the digital twin and causal inference models ought to be tested, helping determine factors such as what is an appropriate set of data to test on.

\paragraph{Establish Data Integrity and Relevance.} The foundation of AI lies in the quality and representativeness of the data on which it is built. As such, validation efforts should assess whether the available data accurately reflects the populations and clinical contexts of the intended application. For digital twins, this will typically mean ensuring access to diverse and comprehensive datasets that include longitudinal measurements. For causal inference, having diverse data is necessary to allow the models to effectively adjust for confounding bias. Ensuring that the data used to train AI models is not incomplete or unrepresentative is paramount to allow the model to generalize well to real-world populations.

\paragraph{Conduct Multifaceted Model Assessment.} Users must acknowledge that model efficacy is multifaceted, and therefore, validation should address a number of important factors, including predictive accuracy, calibration, and robustness. Predictive accuracy measures how correct a model’s predictions are for its designated task. Calibration measures whether model output probabilities align with the true likelihood of the outcomes. Robustness testing evaluates whether models perform reliably under different conditions, such as missing data or the inclusion of outliers. Good performance across these distinct factors is critical for an AI model to perform well in real-world settings, and therefore, validation practices must measure all such aspects of model utility.

\paragraph{Validate Clinical Interpretability.} Interpretability is critical for ensuring that AI models, including digital twins and causal inference methods, are trusted and actionable in clinical settings. Digital twins must provide transparent simulations of patient trajectories that clinicians can understand and validate against their own expertise. For causal inference, interpretability involves ensuring that the reasoning behind any identified causal relationships can be explained.

\paragraph{Benchmark against Traditional Methods.} Benchmarking AI models against existing clinical methods ensures their added value is demonstrable. For example, a digital twin-augmented control arm should be evaluated against traditional placebo arms to confirm its ability to replicate trial outcomes while reducing financial burden. Similarly, the use of causal inference models by clinicians in estimating conditional treatment effects should be examined to determine if they lead to better treatment decisions compared to relying solely on the clinician’s knowledge. Demonstrating that AI-driven methods outperform or complement traditional approaches reinforces their credibility and utility.

\paragraph{Integrate Validation in Real-World Settings.} Validation is incomplete without testing models in real-world clinical scenarios, to assess their ability in their exact intended environment. For digital twins, this might involve deploying the twin alongside a live trial to compare its predictions against observed patient outcomes. For causal models, real-world validation could involve retrospective testing on historical trial data to replicate known findings.

\paragraph{Address Ethical and Regulatory Considerations.} Ethical and regulatory compliance is necessary for AI models in clinical trials. Critically examining the ethics of these models, such as issues related to patient privacy and model bias, is crucial. Transparent documentation of validation procedures is essential to gaining regulatory approval, as is demonstrating that models meet established guidelines for safety and efficacy.

\paragraph{Establish Feedback Loop.} AI models for clinical trials must evolve based on new data and insights, and their evolution should be guided by previous validation results. Continuous feedback loops allow for iterative improvements in accuracy and relevance. For digital twins, this could mean updating simulations as new patient data becomes available, refining predictions over time, while for causal inference, iterative refinement might involve incorporating additional confounders or testing new causal hypotheses as trial data accumulates. Ensuring these models are adaptive, utilizing their validation results for continued improvement, enhances their long-term utility and reliability.

\paragraph{Report Validation Results.} Consistent and transparent reporting of validation processes and outcomes is essential for building trust among clinicians, researchers, and regulators. Such reporting should clearly address how each of the previously outlined points was incorporated in the validation procedure, and what results were achieved by the model. For digital twins, this might include detailed descriptions of how related patient trajectories were collected and learnt from, and how predictions compared to observed outcomes. For causal inference, reports should outline how confounding factors were addressed and how the results align with established causal relationships. Clear, interpretable summaries of validation practices and results ensure stakeholders can assess the reliability and impact of these models on trial efficiency.

This comprehensive validation framework provides a means to ensure that digital twins, causal inference, and other AI models are rigorously tested, allowing for their integration into clinical trials. By addressing an array of aspects in development and deployment, following this framework will ensure that the efficacy, adaptability, ethics, and scientific rigor of AI-driven methodologies are critically assessed, allowing them to ultimately transform the clinical trial landscape.

\begin{table}[ht]
\centering
\renewcommand{\arraystretch}{1.4}
\begin{tabular}{|p{0.95\linewidth}|}
\hline
\textbf{Define the context of use}: Specify model task and application within clinical trials. \\
\hline
\textbf{Establish data integrity and relevance}: Assess data quality, representativeness, and suitability. \\
\hline
\textbf{Conduct multifaceted model assessment}: Evaluate model accuracy, calibration, and robustness across various conditions. \\
\hline
\textbf{Validate clinical interpretability}: Ensure outputs are understandable and actionable for clinicians. \\
\hline
\textbf{Benchmark against traditional methods}: Compare models with existing clinical approaches. \\
\hline
\textbf{Integrate validation in real-world settings}: Test models in live or retrospective trial scenarios. \\
\hline
\textbf{Address ethical and regulatory considerations}: Ensure compliance with ethical and regulatory standards. \\
\hline
\textbf{Establish feedback loop}: Use validation insights to iteratively improve models. \\
\hline
\textbf{Report validation results}: Transparently document and share validation processes and outcomes. \\
\hline
\end{tabular}
\caption{AI model validation framework overview.}
\label{tab:validation-framework}
\end{table}

\section{Additional Ingredients}
While causal inference and digital twins hold transformative potential for clinical trials, their effective integration depends on embedding these technologies within the existing clinical trial ecosystem. By building on the established foundations of clinical research, the additional components we describe below ensure that digital twins and causal models seamlessly augment trial precision, efficiency, and personalization without disrupting existing workflows.

\paragraph{Standardized and Shareable Data Frameworks.} A critical enabler for integrating causal inference and digital twins into clinical trials is the establishment of standardized, dynamic, and shareable data frameworks. These frameworks must ensure that data from diverse sources—such as electronic health records (EHRs), post-marketing surveillance studies datasets, and clinical trial datasets themselves—can be harmonized, annotated, and securely shared while maintaining patient privacy and meeting regulatory requirements. Standardization should include agreed-upon data formats, ontologies, and metadata schemas that facilitate interoperability and reproducibility. For digital twins, dynamically updating datasets are essential to ensure that twins can reflect evolving patient conditions and treatment effects. For causal inference, shared data repositories with rich contextual information enable robust confounding adjustments and the generation of unbiased treatment effect estimates across diverse populations. Collaborative agreements between stakeholders—pharma, healthcare providers, and regulators—are necessary to define data standards and access control, ensuring that AI-driven methods are practically deployable in clinical settings.  

\paragraph{Close Collaboration with Statistical and Trial Methodologists.} Statistical and trial methodologists play a crucial role in ensuring that new technologies align with established statistical frameworks, which are foundational to trial design, analysis, and interpretation. Close collaboration with these experts is therefore important when incorporating AI-based digital twins and causal inference into clinical trials. For digital twins, this means utilising their predictive capabilities within traditional frameworks of hypothesis testing, to ensure that resulting inferences are statistically significant. For example, digital twin-augmented control arms must be rigorously evaluated for their ability to emulate placebo or standard-of-care groups within a trial. Statisticians can help define criteria for evaluating these augmented control arms, ensuring that they produce valid, unbiased, and generalizable results. Collaborating on the creation of validation metrics, such as measures of counterfactual accuracy and calibration, ensures that digital twins enhance rather than disrupt the established statistical paradigms of clinical trials.

For causal inference, statisticians are instrumental in integrating the discussed methods into the broader framework of trial-based causal estimation. For instance, traditional approaches like instrumental variables or propensity score matching can be extended or complemented by AI-driven causal models, which may provide more nuanced adjustments for confounding variables or allow for the evaluation of treatment effects in subpopulations that are difficult to randomize. Statistical methodologists can also contribute to establishing best practices for combining trial data with real-world evidence, harmonizing AI-driven causal models with regulatory requirements for evidence generation. This collaboration is necessary to integrate AI methods into the statistical backbone of clinical trials, enhancing the existing robustness of trial outcomes with the transformative potential of AI.

\paragraph{Interdisciplinary Education and Training.} Education is essential for successfully integrating AI into clinical trials, involving a bidirectional approach to knowledge exchange. Statisticians, trial methodologists, data scientists, clinicians, and regulators must develop a foundational understanding of AI methods, including digital twins and causal inference, to harness their potential effectively \citep{schubert2025ai}. This includes undergoing training in the principles of AI, machine learning techniques, and how they can be applied to clinical trials, such as with recruitment, adaptive trial designs, and real-world data integration. Conversely, AI researchers must gain a deep appreciation of the statistical frameworks that underpin clinical trials, the unique needs and constraints of trial design, and the regulatory requirements governing evidence generation and drug approval. By fostering interdisciplinary education, this effort ensures that AI solutions are both innovative and practical, aligning with the rigorous standards of clinical research while addressing real-world challenges in trial efficiency, safety, and reliability. Collaborative training programs, workshops, and interdisciplinary curricula will be essential in bridging these knowledge gaps and fostering a shared language between these critical stakeholders.

\section{Conclusion: A Call to Action}

This manifesto outlines a shared vision for leveraging AI to transform clinical trials into faster, more personalized, and more inclusive systems. By integrating causal inference and digital twins, we can accelerate the pace of discovery, expand the scope of questions answered, and deliver safer, more effective treatments to patients worldwide. Implementing AI in clinical trials requires a structured approach to ensure success and regulatory compliance. This includes pursuing robust validation, developing best practices for data sharing and standardization, ensuring methods are deployed in a complementary, rather than disruptive fashion, and increasing interdisciplinary education among all stakeholders involved with clinical trials. To scale adoption, pilot projects should target high-impact areas such as oncology or rare diseases. These projects can demonstrate the feasibility of AI-enhanced trials, providing a roadmap for broader implementation. 

The path forward demands collaboration across academia, industry, and regulators. The stakes are high, but the potential rewards—a new era of patient-centered care—are even greater. This is our call to action: to embrace AI as a catalyst for change, redefining clinical trials for the benefit of patients everywhere. The future of medicine depends on it: as we move forward, let us always remember that our actions today will shape the patient care of tomorrow, making it more personalized, more inclusive, and ultimately, more successful.

\section*{Appendix 1: Limiting Pivotal Trial Replication using Digital Twins}

The potential for digital twins to simulate entire clinical trials and predict trial outcomes could make replication of pivotal trials obsolete because they offer a scientifically robust and resource-efficient alternative to traditional trial replication. Below, we discuss how digital twins address the key reasons behind requiring multiple pivotal trials:

\paragraph{Reproducibility Without Physical Replication.} Currently, two pivotal trials are required to replicate results under slightly different conditions and ensure findings are not due to chance or bias. Digital twins, built from comprehensive and dynamic patient and trial data, can simulate entire trial scenarios with high fidelity, including variations in population demographics, treatment protocols, and external factors. By accurately modeling these variations, digital twins can provide robust evidence that mirrors what a second physical trial would achieve, without the need for a second full-scale study. For example, if a first pivotal trial shows significant efficacy, a digital twin could simulate the same trial on an independent population, incorporating variability in factors such as age, comorbidities, or treatment adherence. If the simulated results are consistent with the original trial, they provide additional evidence of reproducibility without requiring a separate physical trial.

\paragraph{Comprehensive Risk-Benefit Assessment.} Digital twins can simulate not only efficacy but also safety outcomes, including rare adverse events that might not appear in a single trial due to limited sample sizes. By using large datasets, digital twins can generate synthetic populations that represent diverse subgroups, including underrepresented populations, allowing for a more comprehensive assessment of a drug’s risk-benefit profile. This eliminates the need for multiple trials to capture rare but critical outcomes. For instance, a digital twin model of a cardiovascular drug could predict safety profiles for patients with multiple comorbidities, such as diabetes and hypertension, addressing concerns that might typically require additional trials.

\paragraph{Testing Variability Across Scenarios.} Regulators require two trials to ensure consistent results across different settings or populations. Digital twins can simulate multiple trial scenarios—e.g., in different geographic locations, healthcare systems, or patient cohorts—providing insights into how the drug might perform under varied conditions. This ability to model population heterogeneity and external influences offers the same validation value as a second trial. For example, in an oncology trial, digital twins can predict how a new immunotherapy would perform in different populations based on genomic data, prior treatment histories, and immune profiles, offering insights that traditionally require additional trials.

\paragraph{Efficiency and Ethical Improvements.} Physical replication of pivotal trials requires significant time, resources, and patient participation. In cases where life-saving drugs are being tested, the delay caused by running two trials can have substantial human costs. Digital twins reduce this delay by offering evidence more rapidly. Furthermore, by reducing the need for placebo-controlled arms in additional trials, digital twins improve the ethical considerations of trials, allowing more participants to receive active treatment. For example, a digital twin can generate a synthetic control arm that mimics the outcomes of a placebo group, enabling the primary trial to focus solely on treatment efficacy without ethical concerns about withholding treatment from control patients.

\paragraph{Dynamic Adaptability.} Digital twins can be updated in real time with new data, making them a continually improving tool for simulation. Unlike a second trial, which provides a static snapshot of results, a digital twin evolves, reflecting real-world outcomes as they emerge. This dynamic capability allows ongoing validation of the initial trial results while accounting for longer-term trends or emerging safety signals.

In conclusion, digital twins could provide a scalable, efficient, and scientifically robust alternative to traditional trial replication. They can simulate diverse trial conditions, assess variability, and ensure reproducibility, addressing the regulatory requirements for multiple pivotal trials. While full regulatory acceptance of digital twins as substitutes for physical trials will require rigorous validation and stakeholder collaboration, their potential to streamline drug approval processes while maintaining or even enhancing scientific rigor makes them a transformative tool for the future of clinical trials.

\bibliography{references}

\end{document}